\documentclass[prd,floatfix,preprintnumbers]{revtex4}

\usepackage{graphicx}
\usepackage{graphicx,epsfig}
\usepackage{bm}
\usepackage{latexsym,amssymb,amsmath,float}
\usepackage{xcolor}

\newcommand {\ga} {\ {\raise-.5ex\hbox{$\buildrel>\over\sim$}}\ }
\newcommand {\la} {\ {\raise-.5ex\hbox{$\buildrel<\over\sim$}}\ }

\def\be{\begin{equation}}
\def\ee{\end{equation}}
\def\ba{\begin{eqnarray}}
\def\ea{\end{eqnarray}}

\begin{document}

\title{Exact general solutions for cosmological scalar field evolution in a
vacuum-energy dominated expansion}
\author{Patrick Hu$^1$ and Robert J. Scherrer$^2$}
\affiliation{$^1$Department of Computer Science, University of Maryland, College Park, MD  ~~20742}
\affiliation{$^2$Department of Physics and Astronomy, Vanderbilt University,
Nashville, TN  ~~37235}

\begin{abstract}
We derive exact general solutions (as opposed to
attractor particular solutions) for the
evolution of a scalar field $\phi$ in a universe
dominated by a background fluid with
equation of state parameter $w_B = -1$, extending earlier work on exact solutions with $w_B > -1$.
Straightfoward exact solutions exist when the evolution is described by a linear differential
equation, corresponding to constant, linear, and quadratic potentials.
In the nonlinear case, exact solutions are derived for $V = V_0\ln \phi$, $V = V_0 \phi^{1/2}$
and $V = V_0/\phi$, and the logarithmic potential also yields an exact first integral. These
complicated parametric solutions are considerably
less useful than those derived previously for a universe dominated by a barotropic fluid such as matter or radiation with $w_B > -1$.  However, we generalize the slow-roll approximation and show that it applies to all sufficiently flat potentials in the case of a vacuum-dominated expansion, while it never applies when the universe is dominated by a background fluid with $w_B > -1$.
\end{abstract}

\maketitle

\section{Introduction}

Scalar fields have long been of interest in cosmology, both as components for models of inflation (see, e.g.,
Refs. \cite{Lyth,Allahverdi} for reviews), and, under the name ``quintessence," as a mechanism to drive
the observed accelerated expansion
of the universe
\cite{RatraPeebles,Wetterich1,Wetterich2,Ferreira1,Ferreira2,CLW,CaldwellDaveSteinhardt,LiddleScherrer,SteinhardtWangZlatev,Wolf}.
(See Refs. \cite{Copeland1,Bamba,Bahamonde,Roy} for reviews).
The equation governing the evolution of a scalar field $\phi$ with potential $V(\phi)$ is
\begin{equation}
\label{phievol}
\ddot{\phi} + 3H\dot{\phi} + \frac{dV}{d{\phi}} = 0,
\end{equation}
where the Hubble parameter $H$ is given by
\begin{equation}
\label{Hdef}
H \equiv \frac{\dot{a}}{a} = \sqrt{\rho/3}.
\end{equation}
In this equation, $a$ is the scale factor, $\rho$ is the total density, and we take $8 \pi G = c = \hbar = 1$ throughout.

In the context of inflation, the appropriate choice for $\rho$ in Eq. (\ref{Hdef}) is the energy
density of the scalar field itself:
\begin{equation}
\label{rhophi}
\rho = \frac{\dot{\phi}^2}{2} + V(\phi). 
\end{equation}
When analyzing quintessence models instead of inflation, we must include both the density of the scalar field as well as the density $\rho_B$ of
any additional background fluid (radiation or nonrelativistic matter).  Denoting the latter by $\rho_B$, we have
\begin{equation}
\label{rhotot}
\rho = \rho_B + \frac{\dot{\phi}^2}{2} + V(\phi).
\end{equation}
In general, Eq. (\ref{phievol}) is intractable when the density is given by Eq. (\ref{rhotot}), although approximate solutions have been derived
for certain conditions on the potential when $\rho_B$ represents nonrelativistic matter \cite{ScherrerSen,ds1,Chiba,ds2,
Andriot}.
The problem can also be inverted to derive potentials corresponding to specific scalar field equations of state
\cite{Dimakis}.   Other approaches to scalar field evolution can be found in Refs. \cite{Alho,Paliathanasis,vanHolten}.

However, in some physically-interesting cases, the contribution of the scalar field to the energy density can be neglected
in comparison to the density of the background component.   For quintessence, this will be the case whenever
the quintessence density is initially subdominant and becomes important only at late times.   
In this case, Eq.
(\ref{phievol}) becomes
\begin{equation}
\label{phiback}
\ddot{\phi} + \frac{2}{1+w_B}\frac{1}{t}\dot{\phi} + \frac{dV}{d{\phi}} = 0.
\end{equation}
This
limiting case, where the background density is usually taken
to be matter or radiation, has been studied extensively.
However, even for this case, most choices of $V(\phi)$ yield a nonlinear
differential equation, and exact solutions cannot be derived.  On the other hand,
particular solutions can often be derived for specific potentials of interest.  These particular solutions lack arbitrary constants and so cannot be
fit to a given set of initial conditions on $\phi$.  However, these solutions can often be shown, under certain conditions,
to act as attractors, so that they describe the asymptotic behavior of $\phi$ for a wide range of initial conditions.
These attractor solutions have been exhaustively studied
\cite{RatraPeebles,Wetterich1,Wetterich2,Ferreira1,Ferreira2,CLW,CaldwellDaveSteinhardt,LiddleScherrer,SteinhardtWangZlatev}.

Ref. \cite{phiexact} addressed a more difficult and less well-examined question:
are there any known exact solutions or first integrals
for Eq. (\ref{phiback})?
When $V$ is constant, linear, or quadratic in $\phi$, Eq. (\ref{phiback}) is a linear differential equation,
and it is easy to find exact solutions for all values of $w_B$.  Further, exact solutions were derived
in Ref. \cite{phiexact} for particular power law potentials in a matter ($w_B = 0$) or radiation ($w_B = 1/3$)
dominated background, and for exponential potentials in a stiff-matter ($w_B = 1$) dominated background.
Note, however, that Eq. (\ref{phiback}) is not valid for the case of a vacuum dominated expansion ($w_B = -1$),
and it is this last remaining case that we explore in this paper.

In the next section, we review the form of Eq. (\ref{phievol}) in a vacuum-dominated background.  In Sec. III, simple solutions
are presented
for constant, linear, and quadratic potentials, for which Eq. (\ref{phievol}) is linear.
In Sec. IV, we derive exact solutions for a handful of power-law and logarithmic potentials, and a first integral
is derived for the logarithmic case.  In Sec. V, we show that the slow-roll approximation can be applied to a wide range
of potentials in the case where the expansion is vacuum-dominated, although it fails when the expansion is dominated by other barotropic fluids such as matter or radiation.
We discuss our results briefly in Sec. VI.

\section{Scalar field equation of motion in a vacuum-dominated background}

We will assume that the expansion of the universe is dominated by a vacuum energy density with $w_B = -1$.
In general, we will take this vacuum energy to be independent of the scalar field itself; for example,
our results will apply in the far future of a $\Lambda$-dominated expansion.  However, these results
also apply when the scalar field energy density is dominant with an equation of state parameter
near $-1$ as, for example, in the case of inflation.

When $w_B = -1$,
the background energy density is constant, giving a constant Hubble parameter $H = H_0$,
and a scale factor that evolves as
\begin{equation}
\label{scalefactor}
a \propto e^{H_0 t}.
\end{equation}
Then
the evolution equation for $\phi$ becomes
\begin{equation}
\label{phivac}
\ddot{\phi} + 3H_0\dot{\phi} + \frac{dV}{d{\phi}} = 0.
\end{equation}
This is the equation for which we seek exact solutions.
These exact solutions will yield two arbitrary constants, which are determined by specifying the values of
$\phi$ and $\dot\phi$ at some fiducial initial time.

Note that Eq. (\ref{phivac}) is qualitatively different from the previously-examined Eq. (\ref{phiback}).
First, Eq. (\ref{phivac}) does not correspond to the $w_B \rightarrow -1$ limit of Eq. (\ref{phiback}).  Second, the first
two terms in Eq. (\ref{phiback}) both scale as the same power of $t$, making it
straightforward to derive particular solutions for power-law and exponential potentials.
This is not the case for Eq. (\ref{phivac}), so
we expect a very different set of exact solutions in the vacuum-energy dominated case.

\section{Constant, linear, and quadratic potentials}

Here we examine potentials of the form
\begin{eqnarray}
\label{linear}
V(\phi) &=& V_0,\\ 
&=& V_1 \phi,\\
&=& V_2 \phi^2.
\end{eqnarray}
where $V_0$, $V_1$, and $V_2$ are constants.  In all of these cases, Eq. (\ref{phiback}) reduces to a linear
differential equation with straightforward solutions.

Consider first the case of a constant potential $V(\phi) = V_0$, so that $dV/d\phi = 0$,
As noted in Ref. \cite{phiexact},
Eq. (\ref{phievol}) with constant $V$
can be solved for arbitrary $H$ to determine $\dot \phi(a)$.  We obtain
\begin{equation}
\dot \phi(a) = C_1 a^{-3},
\end{equation}
where $C_1$ is a constant of integration.  Then the density as a function of the scale factor is just
\begin{equation}
\label{phicon}
\rho_\phi(a) = (C_1^2/2) a^{-6} + V_0,
\end{equation}
i.e., the density evolves as the sum of a constant-density component and a stiff-matter component.  Models of
this sort were dubbed ``skating" models and investigated previously in Refs. \cite{Linder,Sahlen}.
In this case the evolution of $\rho_\phi$ as a function of the scale
factor is the same for both $w_B > -1$ and $w_B = -1$; indeed, Eq. (\ref{phicon}) applies to any functional form for $H$.

Now consider the linear potential $V(\phi) = V_1 \phi$.
In this case, Eq. (\ref{phivac}) yields
\begin{equation}
\phi = C_1 + C_2 e^{-3H_0 t} - \frac{V_1 t}{3H_0}.
\end{equation}
For the background-dominated case with $-1 < w_B < 1$, we have instead \cite{phiexact}
\begin{equation}
\phi = C_1 + C_2 t^{(w_B - 1)/(w_B + 1)} - \left( \frac{1+w_B}{6+2w_B}\right) V_1 t^2.
\end{equation}
Ignoring the transient decaying solutions in both cases, we see that the vacuum-energy dominated evolution of $\phi$
is qualitatively different from the evolution with any other background fluid.
In the the former case, $\phi$ evolves linearly in time, while in all other cases it evolves quadratically.

Now consider the quadratic potential $V(\phi) = V_2 \phi^2$.  In this case,
Eq. (\ref{phivac}) corresponds to the well-known damped harmonic oscillator, with
solution
\begin{equation}
\label{quad1}
\phi = C_{1}e^{b_{+}t} + C_{2}e^{b_{-}t},
\end{equation}
where
\begin{equation}
\label{quad2}
b_{\pm} = \frac{-3H_{0}\pm\sqrt{9H_{0}^{2}-8V_2}}{2}.
\end{equation}
The nature of this solution depends on the sign of $9H_{0}^{2}-8V_2$.
For $9H_{0}^{2}-8V_2 < 0$, the scalar field oscillates in the quadratic
potential well, with an oscillation frequency $\omega = \sqrt{8V_2-9H_{0}^{2}}$
and an amplitude that decays as $e^{-3H_0 t/2} \propto a^{-3/2}$.  Thus,
the energy density scales as $a^{-3}$, which is a general feature of any scalar
field rapidly oscillating in a quadratic potential \cite{Turner}.
However, for the case $9H_{0}^{2}-8V_2 > 0$, both $b_+$ and $b_-$ are real and negative, so the scalar field decays exponentially to zero without oscillating.  In contrast, when $-1 < w_B < 1$, the general solution 
for the quadratic potential is a sum of Bessel functions and is always oscillatory \cite{phiexact}.
	
\section{Power-law and logarithmic potentials}

Now consider more complex potentials, for which Eq. (\ref{phivac}) is nonlinear.
Making the change of variables $x = -3H_0 t$ casts Eq. (\ref{phivac}) into
a standard form that can be found in Ref. \cite{Polyanin}:
\begin{equation}
\label{phivac2}
\phi^{\prime \prime} - \phi^{\prime} = f(\phi),
\end{equation}
with
\begin{equation}
f(\phi) = - \frac{1}{9 H_0^2} \frac{dV}{d{\phi}}. 
\end{equation}
There are at least several dozen forms for $f(\phi)$ that yield exact
solutions, but few of these can be
derived from plausible scalar field potentials.  In particular,
exactly-solvable
forms for $f(\phi)$ with specific numerical coefficients (e.g.,
$f(\phi) = 6\phi + A \phi^{-4}$, where $A$ an arbitrary constant) cannot be considered plausible,
and neither can forms for $f(\phi)$ that require exact
relations between two or more terms, such as
$f(\phi) = A\phi^2 - (9/625)A^{-1}$.  We will therefore make
the simple requirement that $f(\phi)$ correspond to a potential
that can be rescaled by a multiplicative constant.

Applying this plausible constraint eliminates most of the exact solutions
provided in Ref. \cite{Polyanin}, yielding only three potentials
with exact solutions, namely $V = V_0 \ln(\phi)$, $V = V_0 \phi^{1/2}$
and $V = V_0/\phi$, where $V_0$ is a constant.  We consider each of these in
turn.

\subsection{$V = V_0 \ln(\phi)$}
The potential $V(\phi) = V_0 \phi^p (\ln \phi)^q$ was
considered by
Barrow and Parsons \cite{BP} as a model for inflation, and approximate
solutions for the case of a scalar-field dominated expansion  for
various values of $p$ and $q$ were derived.  The potential discussed
here represents the special case where $p=0$ and $q=1$.  Later, Thompson \cite{Thompson}
investigated potentials of the form $V(\phi) = V_0 (\ln \phi)^{\beta}$ as models
for quintessence; the potential considered here corresponds to $\beta = 1$.

For this potential, Eq. (\ref{phivac}) becomes
\begin{equation}
\label{logeq}
\ddot{\phi} + 3H_0\dot{\phi} + \frac{V_0}{\phi} = 0.
\end{equation}
To look for exact solutions, we first determine if there is a first integral
for this equation.  Following Ref. \cite{phiexact}, we make the change of variables
\begin{equation}
t = f(\tau),
\end{equation}
followed by
\begin{equation}
\phi(\tau) = g(\tau) \psi(\tau),
\end{equation}
which transforms Eq. (\ref{logeq}) into
\begin{equation}
\psi^{\prime \prime} + \left[2 \frac{g^{\prime}}{g} + 3 H_0
f^\prime - \frac{f^{\prime \prime}}{f^{\prime}}\right] \psi^\prime
+ \left[ \frac{g^{\prime \prime}}{g} + 3 H_0
f^\prime \frac{g^\prime}{g} - \frac{f^{\prime \prime}}{f^{\prime}}
\frac{g^\prime}{g}\right]\psi + \frac{f^{\prime 2}}{g^2}\frac{V_0}{\psi} = 0,
\end{equation}
where the prime denotes derivative with respect to the new independent variable
$\tau$.  We seek functions $f(\tau)$ and $g(\tau)$ for which the factor multiplying
$\psi^\prime$ is zero, and the factors multiplying $\psi$ and $1/\psi$ are constant.
This can be achieved by taking
\begin{eqnarray}
f(\tau) &=& \frac{1}{3 H_0} \ln \tau,\\
g(\tau) &=& 1/\tau,
\end{eqnarray}
giving
\begin{equation}
\psi^{\prime \prime} + \frac{V_0}{9 H_0^2 \psi} = 0.
\end{equation}
This equation can be integrated to yield
\begin{equation}
\label{psiprime2}
9 H_0^2 \psi^{\prime 2} + {2 V_0} \ln \psi = C,
\end{equation}
where $C$ is a constant.  Transforming back to $\phi$ and $t$, we obtain
\begin{equation}
\label{firstintlog}
\left(\dot \phi + 3 H_0 \phi\right)^2
+ 2 V_0\left(3 H_0 t + \ln \phi\right) = C,
\end{equation}
which provides the first integral for Eq. (\ref{logeq}).  A version
of Eq. (\ref{firstintlog}) is given, without derivation, in the Appendix
of Ref. \cite{Leach}.

To derive the exact solution, we rewrite Eq. (\ref{psiprime2}) as
\begin{equation}
\tau = \int \frac{3 H_0}{\sqrt{C_1 - 2 V_0 \ln \psi}} d\psi + C_2.
\end{equation}
Using our expressions for $t$ and $\phi$ as functions of $\tau$ and $\psi$,
we obtain the parametric solution
\begin{eqnarray}
t &=& \frac{1}{3 H_0} \ln \left[\int \frac{3 H_0}{\sqrt{C_1 - 2 V_0 \ln \psi}} d\psi + C_2 \right],\\
\phi &=& \psi \left[\int \frac{3 H_0}{\sqrt{C_1 - 2 V_0 \ln \psi}} d\psi + C_2 \right]^{-1}.
\end{eqnarray}
A different parametric solution for this case can be found in Ref. \cite{Polyanin}.  As expected, the
solution contains two arbitrary constants, which are fixed by the values of $\phi$ and $\dot \phi$ at
a given initial time $t$.  A phase diagram illustrating the evolution in the $(\phi, \dot \phi)$ plane is given in the
next section.

\subsection{$V(\phi) = V_0 \phi^{1/2}$}
Now consider the case $V(\phi) = V_0 \phi^{1/2}$, for which Eq. (\ref{phivac}) takes
the form
\begin{equation}
\label{phihalf}
\ddot{\phi} + 3H_0\dot{\phi} + \frac{1}{2}{V_0}\phi^{-1/2} = 0.
\end{equation}
We define a new independent variable $u$ given by
\begin{equation}
\label{usub}
u = \dot \phi + 3H_0 \phi,
\end{equation}
and take the dependent variable to be
\begin{equation}
V = V_0 \phi^{1/2}.  
\end{equation}
These substitutions are derived from a series of procedures outlined in Ref. \cite{Kamke}.
Since $dV/du = \dot V/\dot u$, Eq. (\ref{phihalf}) corresponds to
\begin{equation}
\frac{dV}{du} + u = 3H_0 \left(\frac{V}{V_0}\right)^{2},
\end{equation}
which is a Riccati equation.
This equation can be transformed into a linear second-order equation through the
standard substitution
\begin{equation}
\label{wsub}
V = - \frac{V_0^2}{3H_0}\frac{1}{w}\frac{dw}{du},
\end{equation}
where $w$ is the new independent variable.  This substitution gives
\begin{equation}
\frac{d^2 w}{du^2} - \frac{3H_0}{V_0^2} uw = 0.
\end{equation}
This is the Airy equation, with solution
\begin{equation}
\label{Airy}
w = C_1 Ai\left((3H_0/V_0^2)^{1/3} u  \right) + C_2 Bi\left((3H_0/V_0^2)^{1/3}
u\right)
\end{equation}
where $Ai$ and $Bi$ are the Airy functions.  We now invert our sequence of
substitutions to give the exact (albeit parametric) solution to Eq. (\ref{phihalf}).
A similarly complex parametric solution is given in
Ref. \cite{Polyanin}.

\subsection{$V = V_0/\phi$}
Finally we consider $V = V_0/\phi$, for which Eq. (\ref{phivac}) takes
the form
\begin{equation}
\label{phitwo}
\ddot{\phi} + 3H_0\dot{\phi} - \frac{V_0}{\phi^{2}} = 0.
\end{equation}
We take $u$ to be given by Eq. (\ref{usub}), but now $u$ will be the dependent variable
instead of the independent variable.  The new independent variable will be
\begin{equation}
v = \frac{1}{2} (\dot \phi + 3 H_0 \phi)^2 + \frac{V_0}{\phi}.
\end{equation}
Again, these substitutions are derived from a sequence of procedures taken from Ref. \cite{Kamke}.
Then we have
\begin{equation}
\frac{du}{dv} = \frac{\dot u}{\dot v} = \frac{1}{3 H_0 V_0}\left( -\frac{1}{2} u^2 +v \right).
\end{equation}
This is again a form of the Riccati equation, and we make the substitution
\begin{equation}
u =  6 H_0 V_0 \frac{1}{w}\frac{dw}{dv},
\end{equation}
giving another Airy equation:
\begin{equation}
\frac{d^2 w}{dv^2} - \frac{1}{18 H_0^2 V_0^2} vw = 0,
\end{equation}
with solution
\begin{equation}
\label{Airy2}
w = C_1 Ai\left((1/18 H_0^2 V_0^2)^{1/3} v  \right) + C_2 Bi\left((1/18 H_0^2 V_0^2)^{1/3}
v\right).
\end{equation}
As in the previous case, this sequence can be reversed to derive a parametric solution, a version of which
can also be found in Ref. \cite{Polyanin}.

\section{Generalizing the slow-roll condition}

In models of inflation, one often invokes the slow-roll condition to derive approximate solutions to Eq. (\ref{phievol})
for the case where the scalar field dominates the expansion (see, e.g., Refs. \cite{LPB,LiddleLyth}).
This approximation assumes that the first term in Eq. (\ref{phievol}) is negligible compared to the other two terms,
so that one can write
\begin{equation}
\label{slowrollinf}
3 H \dot \phi = - V^\prime(\phi),
\end{equation}
which requires
\begin{equation}
\label{slowinf1}
(V^\prime/V)^2 \ll 1,
\end{equation}
and
\begin{equation}
\label{slowinf2}
|V^{\prime \prime}/V| \ll 1,
\end{equation}
where the prime denotes the derivative with respect to $\phi$.

However, the inflationary slow-roll condition is just a special case of the method of dominant balance, which applies whenever one or more terms in a differential equation are subdominant.  Here we generalize the inflationary slow-roll condition to arbitary $H$ and show how it can be applied to the case under consideration in this paper.
We begin by determining the conditions for which the first term in Eq. (\ref{phievol}) is dominated by the other two, so that Eq. (\ref{slowrollinf}) applies.
Taking the time derivative of Eq. (\ref{slowrollinf}) and requiring
that $\ddot \phi \ll V^\prime(\phi)$ gives
the condition
\begin{equation}
\label{phibalance}
\frac{V^{\prime \prime}}{9 H^2} + \frac{\dot H}{3 H^2} \ll 1,
\end{equation}
which can be reexpressed as
\begin{equation}
\label{slowrollgen}
\frac{1}{6} \frac{V^{\prime\prime}}{V} \Omega_\phi (1-w_\phi) - \frac{1}{2}(1+w_T) \ll 1.
\end{equation}
Here $w_\phi$ and $w_T$ are the equation of state parameters for the scalar field and the total energy content of the Universe, respectively.
Barring an accidental cancellation between the two terms,
this inequality requires
\begin{equation}
\label{slowroll1}
1+w_T \ll 1,
\end{equation}
and
\begin{equation}
\label{slowroll2}
\frac{V^{\prime\prime}}{V} \Omega_\phi (1-w_\phi) \ll 1.
\end{equation}
In the case of inflation, the scalar field is the only component, and the flatness of the potential ensures that
$w_\phi = w_T \ll 1$, while the second slow-roll condition (Eq. \ref{slowinf2}) ensures that Eq. (\ref{slowroll2}) is satisfied.  On the other hand,
when the universe is dominated by a barotropic background fluid with
$w_T > -1$, (as, e.g., in Refs. \cite{LiddleScherrer} or \cite{phiexact}), Eq. (\ref{slowroll1}) is not
satisfied, so the slow-roll approximation will never apply. As a corollary, the slow-roll approximation never applies to quintessence models in which the universe is initially matter-dominated and evolves to the present state with $\Omega_\phi \approx 0.7$, as $w_T$ is never close to $-1$ (a point emphasized by Linder \cite{Linderpaths}).

However, the slow-roll approximation {\it does} apply to some of the models under consideration here.  By construction, an assumed constant value of $H$ corresponds exactly to $w_T = -1$, so condition (\ref{slowroll1}) is automatically satisfied.  Then we can rewrite condition (\ref{slowroll2}) in terms of the parameters in Eq. (\ref{phibalance}) as
\begin{equation}
\label{slowroll3}
V^{\prime \prime} \ll H_0^2.
\end{equation}
Any sufficiently flat potential, in the sense of Eq. (\ref{slowroll3}), will
lead to slow-roll evolution that is well-approximated
by $3 H_0 \dot \phi = - V^\prime(\phi)$, so that $\phi$ is given by
\begin{equation}
\label{slowsolution}
\int\frac{d\phi}{V^\prime(\phi)} = C -\frac{t}{3H_0},
\end{equation}
where $C$ is a constant determined by initial conditions.

When Eq. (\ref{slowroll3}) applies, the evolution given by
Eq. (\ref{slowsolution}) this will be an excellent approximation.  Consider first the quadratic potential
examined in Sec. III.  The slow-roll condition is satisfied in this case as long as $V_2 \ll H_0^2$, in which case the slow-roll solution from Eq. (\ref{slowsolution}) is
\begin{equation}
\phi = Ce^{-2 V_2 t/3H_0}.
\end{equation}
This is exactly the limiting case of the exact solution (Eqs. \ref{quad1}$-$\ref{quad2}) for $V_2 \ll H_0^2$, as expected.

For the case of a positive power law potential, $V(\phi)  = V_n \phi^n$, with $n > 2$, the slow roll condition corresponds to
$\phi \ll (H_0^2/V_n)^{1/(n-2)}$, so for sufficiently small $\phi$, the asymptotic evolution will be given by
\begin{equation}
\label{slowroll4}
\phi = \left[C + n(n-2)(V_n/3H_0)t\right]^{-1/(n-2)}.
\end{equation}
Thus, for $n>2$, the scalar field always evolves smoothly to zero without oscillation.  Conversely,
for power law models with $n < 0$, the slow-roll condition gives a lower bound on $\phi$, namely $\phi \gg (V_n/H_0^2)^{1/(2-n)}$, and when this condition is satisfied we again obtain Eq. (\ref{slowroll4}), except
that now this equation represents asymptotic evolution toward arbitrarily large
$\phi$.  Note however, that these arguments fail for power-law law models with
$0 < n < 2$.  Such models evolve asymptotically to $\phi = 0$, so that $V^{\prime \prime} \rightarrow \infty$.  For these models the slow-roll condition can be satisfied at early times if $\phi$ is sufficiently large, but it fails in the long-time asymptotic limit.

Finally, consider the logarithmic potential from the previous section.  The slow-roll condition fails for this potential as $\phi \rightarrow 0$, but it holds for sufficiently large values of $\phi$.  In that case,
Eq. (\ref{slowsolution}) gives
\begin{equation}
\label{slowrolllog}
\phi = \sqrt{C - 2 V_0 t/3 H_0}.
\end{equation}
This result can be compared with the exact first integral for this potential, Eq. (\ref{firstintlog}).  In the slow-roll regime, $\dot \phi \ll 3H_0 \phi$ and $3 H_0 t \gg \ln \phi$, so that Eq. (\ref{firstintlog})
reduces to Eq. (\ref{slowrolllog}).  Of course, the first integral in Eq. (\ref{firstintlog}) is valid at all times,
including in the limit where $\phi \rightarrow 0$, at which point Eq. (\ref{slowrolllog}) fails.

When Eqs. (\ref{logeq}) and (\ref{firstintlog}) are rewritten in terms of $(H_0/\sqrt{V_0}) \phi$ instead of $\phi$ and $H_0 t$ instead of $t$, the evolution becomes independent of $V_0$ and $H_0$ and depends
only on the initial value of $\phi$. 
The phase-space evolution for the logarithmic potential in terms of these variables is shown in Fig. 1 for $\dot \phi = 0$ at $t=0$.
In this phase plane, the slow-roll attractor (Eq. \ref{slowrolllog}) becomes
\begin{equation}
[(H_0/\sqrt{V_0}) \phi][\dot \phi/\sqrt{V_0}] = -1/3,
\end{equation}
which is shown in Fig. 1 as a dashed curve.  As expected, this is an excellent approximation for large $\phi$ but becomes increasingly inaccurate as $\phi$ decreases.  Note, however, that
all of the curves converge to identical behavior at late times, distinct from the slow-roll prediction.  This exact
attractor can
be understood in terms of the first integral in Eq. (\ref{firstintlog}), which can be rewritten in terms of $\phi_0$,
the initial value for $\phi$ at $t=0$, assuming $\dot \phi = 0$ at $t=0$:
\begin{equation}
\left(\dot \phi + 3 H_0 \phi\right)^2
+ 2 V_0 \ln \phi  = 9 H_0^2 \phi_0^2
+ 2 V_0 \ln \phi_0 - 6 V_0 H_0 t.
\end{equation}
Clearly, a change in $\phi_0$ can be compensated by a time translation on the right-hand side of the
equation, yielding an identical trajectory in the $(\phi, \dot \phi)$ plane, as is apparent in Fig. 1.

\begin{figure}
\centering
\includegraphics[width=7in]{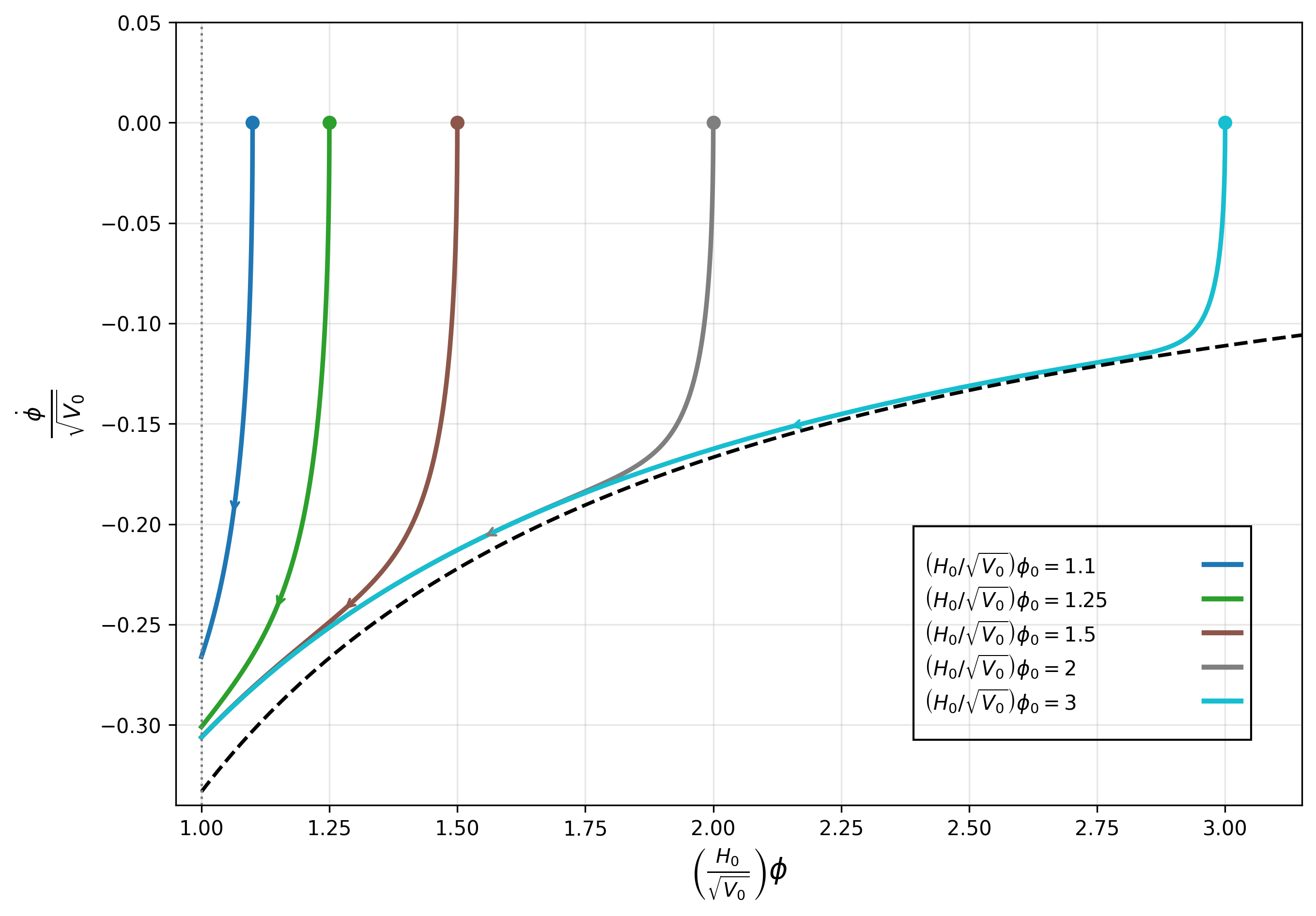}
\caption{Evolution in the $(\phi, \dot \phi)$ plane as a function of time for the potential $V = V_0 \ln(\phi)$ with the indicated initial values of $\phi$,
and $\dot \phi = 0$ at $t=0$.  Dashed curve is the slow-roll attractor.}
\end{figure}

\section{Applications}

As noted in Sec. II, the results derived here are applicable to both quintessence (in the case where the universe contains both a dominant cosmological constant and a subdominant scalar field) and to inflation (during the epoch when the equation of state parameter of the scalar field is close to $-1$).

As an example, consider a universe with a cosmological constant and a scalar field with a quadratic potential as in Sec. III, $V(\phi) = V_2 \phi^2$, and
assume that the scalar field is subdominant.  In that case the evolution
of the scalar field is given by Eqs. (\ref{quad1}) and (\ref{quad2}), with
$b_+$ and $b_-$ both real and negative.  If the field is initially frozen with $\dot \phi = 0$ and $\phi = \phi_0$ at $t=0$, we can write the solution as
\begin{equation}
\phi = \phi_0 e^{-3 H_0 t/2}[\cosh(\beta t) + (3H_0/2\beta) \sinh(\beta t)],
\end{equation}
where $\beta$ is given by
\begin{equation}
\beta = \frac{\sqrt{9 H_0^2 - 8 V_2}}{2}.
\end{equation}
Using this expression for $\phi(t)$, the equation of state parameter for the quintessence field is
\begin{equation}
\label{wquad}
w = \frac{2 V_2 - (\beta \coth(\beta t) + 3H_0/2)^2}
{2 V_2 + (\beta \coth(\beta t) + 3H_0/2)^2}
\end{equation}
It is easy to see from this expression that the quintessence field begins
at $t=0$ with $w=-1$ (since $\dot \phi = 0$), and it evolves asymptotically to 
\begin{equation}
w = - \frac{2 \beta}{3 H_0}.
\end{equation}

A more plausible scenario arises if the scalar field itself corresponds to the presently-observed dark energy, without an additional background cosmological constant.  In this case, $w$ for the quintessence field will be $-1$ initially but will increase at late times as the field
rolls downhill, corresponding to a classic ``thawing" scalar field scenario \cite{CaldwellLinder}.  Although our results in this case are
no longer exact, they provide an excellent approximation as long as Eq. (\ref{scalefactor}) is close to the true evolution of the scale factor. This will be the case
as long as $w$ is not too far from $-1$.  In fact, we find that Eq. (\ref{wquad}) gives a very accurate (better than 5\%) approximation
to the true value of $w$ for $-1 < w < -0.65$.

\section{Discussion}

This paper completes the project initiated in Ref. \cite{phiexact} to determine all exact solutions for the evolution of a scalar field in an expanding universe with a barotropic fluid background. 
These exact solutions provide a more complete description of the scalar field evolution than any corresponding attractor solutions.  An attractor represents only an asymptotic solution valid for a given range of initial conditions or parameter values.  By definition, the exact solutions we have derived here will always agree asymptotically with any corresponding attractor solutions.

As noted, the case of vacuum energy domination is unique and results in a very different differential equation from other barotropic solutions, and
the resulting set of solutions is considerably less useful than those derived in Ref. \cite{phiexact}.  While exact parametric solutions exist for $V = V_0\ln \phi$, $V = V_0 \phi^{1/2}$ and $V = V_0/\phi$, the last two of these are so complex that they are unlikely to be of much practical value.  On the other hand, the logarithmic potential yields both a simpler parametric solution and, more importantly, an exact first integral.  As in the case of the potentials discussed in Ref. \cite{phiexact}, the first integral is actually of greater utility than the exact solution itself, as it demarcates the scalar field evolution in terms of the phase space, $(\phi, \dot \phi)$, as a function of time.

On the other hand, we have shown that the (non-exact) slow roll approximation can be applied to almost all models examined here, as long as $V^{\prime \prime}$ is sufficiently small.  This contrasts with scalar field evolution in a universe  dominated by a background barotropic fluid with equation of state parameter $w_B > -1$, for which the slow-roll approximation {\it never} applies.  In that sense, the models under consideration here can be ``solved" more easily than those examined in Ref. \cite{phiexact}.




\end{document}